\documentstyle[12pt,russian]{article}
\def\l{\lambda}
\def\s{\sigma}
\def\ra{\rightarrow}

\def\bb{\begin{equation}}
\def\ee{\end{equation}}

\def\im{\mbox{Im}}
\begin{document}

{\large

\title{The KdV equation on a half-line}
\bigskip
\bigskip
\author{I.T.Habibullin\thanks{Corresponding author,
e-mail: ihabib@imat.rb.ru} and A.N.Vil'danov}
\date{\null}
\maketitle
Sterlitamak State Pedagogical Institute: Lenin str., 37,
Sterlitamak, 453103, Russia
\vspace{0.5cm}

Abstract. The initial boundary value problem on a half-line for
the KdV equation with the boundary conditions
$u\vert_{x=0}=a\leq0$, $u_{xx}\vert_{x=0}=3a^2$ is
integrated by means of the inverse scattering method.
In order to find the time evolution of the scattering
matrix it turned out to be sufficient to solve
the Riemann problem on a hyperelliptic curve of genus two,
where the conjugation matrices are effectively defined by
initial data.
\vspace{0.5cm}

Mathematical Subject Classification (1991): 35 and 46.

Key words: integrability, scattering matrix, Riemann-Hilbert problem.


\section{Introduction}
Consider the initial boundary value problem for the KdV
equation on the first quadrant
\begin{eqnarray}&& u_t=u_{xxx}-6uu_x,\quad x>0,\,\,t>0,
                          \label{kdv}\\
&&u\vert_{x=0}=a,\quad u_{xx}\vert_{x=0}=b,
                          \label{x=0}\\
&&u\vert_{t=0}=u_0(x),\quad u_0(x)\vert_{x\ra+\infty}\ra0.
                          \label{t=0}
\end{eqnarray}
The initial value $u_0(x)$ is supposed to be consistent
with the boundary condition (\ref{x=0}) at the corner
point: $u_0(0)=a$, $u_{0xx}(0)=b$, the boundary data $a$
and $b$ are chosen to be constants. Besides $u_0(x)$
should be a smooth function decreasing rapidly enough.
Remind that the problem (\ref{kdv})-(\ref{t=0}) is
correctly posed and uniquely soluble (see, for instance, \cite{ton}).

The problem (\ref{kdv})-(\ref{t=0}) admits
infinitely many integrals of motion. The first three of them
are of the form
\begin{eqnarray}
&&\int^{\infty}_{0}udx=(3a^2-b)t+\mbox{const},
\nonumber \\
&&\int^{\infty}_{0}(u^2_x+2u^3)dx=-(3a^2-b)^2t+\mbox{const},
\nonumber\\
&&\int^{\infty}_{0}(u^2_{xx}-5u^2u_{xx}+5u^4+
4(b-3a^2)u^2)dx=\mbox{const}.
\nonumber
\end{eqnarray}
When the parameter $3a^2-b$ is different from zero then evidently the first
two integrals of motion contain the explicit $t$-dependence. It can be proved
(see \cite{aggh}) that the problem (\ref{kdv})-(\ref{t=0}) is consistent with
an infinite number of higher symmetries and, because of this reason, the KdV
equation (\ref{kdv}) admits a large class of explicit algebra-geometric
solutions satisfying the boundary condition (\ref{x=0}) (see \cite{ahs}). All
these facts allow one to hope that the problem (\ref{kdv})-(\ref{t=0}) can be
integrated by means of the inverse scattering method. The scattering matrix
$s(\xi,t)$ of the associated linear operator $y''=(u(x,t)-\lambda)y$, $x>0$,
defined in the standard way evaluates in time $t$ by means of the following
system of equations
\bb
s_t=4i\xi^3[s,\s_3]+(u_x\s_1-{4u\xi^2-u_{xx}+2u^2\over
2\xi}\s_2+ {2u^2-u_{xx}\over2\xi}i\s_3)s
\label{st}
\ee
containing unknown $u_x(0,t)$ and given $u(0,t)=a$ and
$u_{xx}(0,t) =b$ (here and below $\s_j$ are the Pauli
matrices). The equation  (\ref{st}) shows that
the nonlocal change of variables from
$u(x,t)$ to $s(\xi,t)$ doesn't lead to any separation of
variables in the problem  (\ref{kdv})-(\ref{t=0}). Note
that such kind obstacles always arise when the initial
boundary value problem is studied for integrable equations
(see, for instance, \cite{fi,f} where an alternative approach
to the problem is discussed).

The system (\ref{st}) is under\-de\-ter\-mi\-ned, both the coefficient and the
solution are not given (that is why the system is really nonlinear). But it is
worth to note that in addition to the system there is an extra condition: for
all values of $t$ the solution $s(\xi,t)$ has to preserve its analytical
properties, i.e. has to belong to the class of scattering matrices. The latter
allows one in some particular cases to simplify essentially the system
(\ref{st}) and ever reduce it to a problem of factorisation. Let us do first a
linear change of dependent variables by setting $s(\xi,t)=T(\xi)S(\xi,t)\exp
(4i\xi^3\s_3t)$. Then the new variable $S(\xi,t)$ solves the equation
\bb
S_t=(i\nu\s_3+u_x\s_1)S,
\label{St}
\ee
where
$$ T(\xi)=\pmatrix{q&q\cr-i\nu&i\nu},\quad q=4\xi^2+2a,$$
and
\bb
\nu(\xi)=\sqrt{16\xi^6+4\xi^2(b-3a^2)+2ab-4a^3}. \label{nu}
\ee

Suppose the function $P=P(\xi)$ to define an involution
of the curve (\ref{nu}) such that $\nu(\xi)=\nu(P(\xi))$
for all $\xi$. Then evidently the identity
\bb
S^{-1}(P(\xi),t)S(\xi,t)=S^{-1}(P(\xi),0)S(\xi,0)
\label{id}
\ee
holds, which may be considered as the first integral of the system (\ref{St}).
The problem arises how to solve the nonlinear equation (\ref{id}) for $t>0$.
Below in the paper we discuss the particular case when the additional
constraint $b=3a^2$, $a\leq0$ is imposed and however we act in a slightly
different way (the homogeneous case $a=b=0$ has been studied in \cite{h1999}).
Here we find the solution of the t-equation at the straight-line $x=0$ by
gluing up the eigenfunctions of the two operators from the Lax pair and then
solving the Riemann problem of linear conjugation for piecewise analytic
functions on a six sheeted Riemann surface. The main observation we use below
is as the $x$- and $y$-equations have not only a common fundamental solution
but a common solution of the scattering problem for all values of the spectral
parameter except some contour on the complex plane.

\section{The direct scattering problem}

It is well known that the KdV equation is the consistency
condition of the following two systems of ordinary linear
differen\-tial equations:
\begin{eqnarray}
&&Y_x=UY,                                \label{yx}\\
&&Y_t=VY,
                                     \label{yt}
\end{eqnarray}
where the coefficient matrices are
$$
U=\pmatrix{0&1\cr u-\l&0},
\quad
V=\pmatrix{u_x&-4\l-2u\cr u_{xx}-(4\l+2u)(u-\l)&-u_x}.$$
Due to the boundary conditions (\ref{x=0}) the equation
(\ref{yt}) takes the form
\bb
Y_t=\pmatrix{u_x&-4\l-2a\cr b-(4\l+2a)(a-\l)&-u_x}Y
\label{yt0}
\ee
along the border $x=0$. In other words the linear systems
(\ref{yx}), (\ref{yt}), and (\ref{yt0}) give the Lax
representation of the initial boundary value problem
(\ref{kdv}), (\ref{x=0}), and (\ref{t=0}).

Define matrix-valued solutions of the auxiliary system (\ref{yx}), satisfying
the following asymptotic for all real values of the parameter $\xi=\sqrt\l$
\begin{eqnarray}
&&Y_1(x,t,\xi) \ra T_0(\xi) e^{ix\xi\sigma_3}, \quad x\ra
+\infty,     \label{y1}\\
&&Y_2(x,t,\xi) \ra T_0(\xi), \qquad\quad x\ra
0,
\label{y2}
\end{eqnarray}
where $T_0(\xi)=\pmatrix{1&1\cr i\xi&-i\xi}.$ The columns $\psi_k(x,t,\xi)$
and $\phi_k(x,t,\xi),$ $k=1,2$ of the matrices $Y_1=(\psi_1,\phi_1)$ and
$Y_2=(\phi_2,\psi_2)$ are known to have analytical continuations with respect
to the parameter $\xi$ from the real axis onto the complex plane. Moreover,
$Y_2(x,t,\xi)$ is an entire analytical function and its columns have the
following asymptotic in the corresponding half-planes when $\vert\xi\vert$
goes to infinity
\begin{eqnarray} &\phi_2(x,t,\xi) \ra
e^{ix\xi}\pmatrix{1\cr i\xi}(1+O(\xi^{-1})),&\quad
\im\xi\leq0,\label{phi2}\\ &\psi_2(x,t,\xi)\ra
e^{-ix\xi}\pmatrix{1\cr
-i\xi}(1+O(\xi^{-1})),&\quad \im\xi\geq0. \label{psi2}
\end{eqnarray}

In general the columns of the other solution $Y_1$ are defined only on the
half-planes $\im\xi<0$ and $\im\xi>0$, respectively, where they have similar
asymptotic for $\xi\ra\infty$
\begin{eqnarray}
&\phi_1(x,t,\xi)\ra e^{-ix\xi}\pmatrix{1\cr
-i\xi}(1+O(\xi^{-1})),&\quad \im\xi\leq0,\label{phi1}\\
&\psi_1(x,t,\xi)\ra e^{ix\xi}\pmatrix{1\cr
i\xi}(1+O(\xi^{-1})),&\quad \im\xi\geq0.\label{psi1}
\end{eqnarray}
The scattering matrix of the system (\ref{yx}) given on
the half-line is defined as a ratio of two fundamental
solutions: $s(\xi)=Y_2^{-1}Y_1$. It is easy to show that
entries $s_{11}$ and $s_{21}$ of the matrix $s$ are
analytic in the domain $\im \xi>0$ and similarly the
entries $s_{12}$ and $s_{22}$ -- in the domain $\im
\xi<0$.

Let us compound matrices $\phi$ and $\psi$ by taking the
vectors $\phi_i$ and $\psi_i$ as their columns multiplied
by scalar factors as follows
\begin{eqnarray}
&\phi(x,\xi)=(\phi_2,\phi_1s^{-1}_{22})(x,\xi) e^{-ix\xi\s_3},
\label{phi}\\
&\psi(x,\xi)=(\psi_1s^{-1}_{11},\psi_2)(x,\xi) e^{-ix\xi\s_3}.
\label{psi}
\end{eqnarray}

By definition the functions $\phi(x,\xi)$ and
$\psi(x,\xi)$ are analytic in lower and upper half-planes,
respectively. At the real axis $\im\xi=0$ they are related
to each other by conjugation condition:
\bb
\phi(x,\xi)=\psi(x,\xi)r(x,\xi).           \label{rp1}
\ee
The $x$ dependence  of the conjugation matrix
$r(x,\xi)$ is given for $x\geq0$ by:
\bb
r(x,\xi)=e^{ix\xi\s_3}r(0,\xi)e^{-ix\xi\s_3}.
                        \label{rx}
\ee
The asymptotic representations above yield
\begin{eqnarray}
&\phi(x,\xi)\ra T_0(\xi)\quad \mbox{for}\quad \im
\xi\leq0,\,
\xi\ra
\infty & \mbox{and}               \label{phia}\\
&\psi(x,\xi)\ra T_0(\xi)\quad
\mbox{for}\quad \im \xi\geq0,\, \xi\ra \infty.&
                        \label{psia}
\end{eqnarray}
The eigenfunctions (\ref{phi}), (\ref{psi}) are degenerate at the
point $\xi=0$, since the scattering matrix has generally a pole there.
Evidently,
\bb
\phi(x,0)\pmatrix{1\cr0}=0,
\quad
\psi(x,0)\pmatrix{0\cr1}=0.       \label{zero}
\ee
In the special case when $s(\xi)$ is bounded at $\xi=0$
the condition (\ref{zero}) is not valid.

The conjugation matrix in (\ref{rp1}) is expressed
explicitly through entries of the scat\-te\-ring matrix:
\bb
r(x,\xi)=\pmatrix{1&s_{12}s_{22}^{-1}e^{2ix\xi}\cr
-s_{21}s_{11}^{-1}e^{-2ix\xi}&s_{11}^{-1}s_{22}^{-1}}.             \label{r}
\ee

A non-degenerate matrix-valued solution $Y(x,\xi)$ of the
system (\ref{yx}) is called solution of the scattering
problem on the half-line $x\geq0$ at the point $\xi$,
$\im\xi\neq0$, if the function $Y(x,\xi)e^{-ix\xi\s_3}$ is
bounded uniformly for all values of $x\geq0$. Clearly
functions $Y(x,\xi)=\phi(x,\xi)e^{ix\xi
\s_3}$ and $Y(x,\xi)=\psi(x,\xi)e^{ix\xi \s_3}$ are solutions
of the scattering problem at the corresponding half-planes
except the points $\xi$ where $s_{22}(\xi)=0,$ if
$\im\xi<0$ and $s_{11}(\xi)=0,$ if $\im\xi>0$.

{\bf{Remark.}} The solution of the scattering problem on a
half-line is not unique. It is defined up to the right
side matrix-valued multiplier upper triangular when
$\im\xi>0$ and lower triangular when $\im\xi<0$.

We will use this freedom when finding solutions of the
similar scattering problem for the second operator which
is responsible for the $t$-evolution.

Let us give explicit representations of the matrices
$\phi$ and $\psi$ at the end $x=0$
\begin{eqnarray}
T_0^{-1}(\xi)\phi(0,\xi)=\pmatrix{1&{s_{12}(\xi)\over s_{22}
(\xi)}\cr 0&1},
 \, T_0^{-1}(\xi)\psi(0,\xi)=\pmatrix{1&0\cr
{s_{21}(\xi)\over s_{11}(\xi)}&1}.
                           \label{Tpsi}
\end{eqnarray}


\section{Method of gluing up the eigenfunctions}

In the previous section we constructed eigenfunctions
(i.e. solutions of the scattering problem) for the
$x$-equation (\ref{yx}). According to our conjecture there
should be a common eigenfunction for both $x$- and
$t$-equations (\ref{yx}) and (\ref{yt0})  at the corner
point ($x=0$, $t=0$) defined for all values of the spectral
parameter. But how to find such a solution? The matter is
that the scattering problem for these two equations are
naturally formulated on two different spectral planes.
Really, after the linear change of variables
$Y(t,\xi)=T(\xi)\tilde Y(t,\xi)$ (\ref{yt0}) takes a form
of the well-known Zakharov-Shabat system (cf. (\ref{St}))
\bb
\tilde {Y}_t=(i\nu\s_3+u_x(0,t)\s_1)\tilde {Y},
\label{ytn}\ee
where the function
\bb\nu=\sqrt{16\xi^6+2a^3}
\label{g}\ee
is a particular case of (\ref{nu}) under the constraint
$b=3a^2$.

To be able to compare functions on $\xi$ with those on $\nu$ it is necessary
to consider two Riemann surfaces $\Gamma$ and $H$ appearing in a natural way
such that the function (\ref{g}) establishes a one-to-one correspondence
between them. The first one is a two-sheeted cover of the $\xi$-plane glued by
its six cuts such that two cuts go along the real axis four others are done to
make surface unvariant under rotation by angle ${\pi\over3}$ (see Pic.1).
Denote through $\Gamma_1$ and $\Gamma_2$ the sheets of the surface. Require
that $\xi\in\Gamma_1$ if the asymptotic representation $\nu (\xi)=4\xi^3
+o(1)$ is valid for $\xi
\rightarrow \infty$ and $0<\mbox{arg}\xi<{\pi\over3}$.
Remind that the continuous spectrum of the equation
(\ref{yx}) coincides with the real axis $\im \xi=0$. By
rotating the axis around the origin by angles
${\pi\over3}$ and ${2\pi\over3}$, i.e. applying the
involution of the curve, one gets two more
directed lines. The conjugation contour $L$, consisting
of these three lines taken on both sheets,
divides the surface into twelve sectors. Enumerate them in
the following way. Denote through $I_{1,1}$ and $I_{2,1}$
those lying in the first quadrant on the sheets $\Gamma_1$
and $\Gamma_2$ respectively. Denote $I_{2,1}$ and
$I_{2,2}$ the next two sectors that lie partly in the
first and partly in the second quadrants on the
corresponding sheets and so on.

The function $\xi=\xi(\nu)$ inverse to (\ref{g}) is evidently single-valued on
the six-sheeted cover of the complex plane $H$ with sheets $H_j$, $j=1,...,6$
glued by two cuts lying on the imaginary axis and going to infinity from the
branch points $\nu_{\pm}=\pm i\sqrt{2\vert a\vert^3}$ (see Pic.2). The sheets
are enumerated in such a way that the images $\tilde I_{k,j}=\nu(I_{k,j})$ are
as follows
\begin{eqnarray}
&\tilde I_{1,k}=\{\nu\in H_k, \mbox{Re}\nu>0\}\cup
\{\nu\in H_{k+1}, \mbox{Re}\nu<0\}\cap\{\mbox{Im}\nu>0\},&\nonumber\\
&\tilde I_{2,k}=\{\nu\in H_k, \mbox{Re}\nu<0\}\cup
\{\nu\in H_{k+1}, \mbox{Re}\nu>0\}\cap\{\mbox{Im}\nu<0\}.&\nonumber
\end{eqnarray}
The sheets are such that when crossing the upper cut from the left
to the right one passes from $H_j$ to $H_{j+1}$, where
$H_7=H_1$ and $H_{0}=H_{6}$.

Denote through $Y_{kj}(t,\nu)=\Psi_{kj}(t,\nu)e^{-4i\xi^3\sigma_3}$ solutions
of the equation (\ref{ytn}) where the multipliers $\Psi_{kj}(t,\nu)$  are
regular for large enough $\nu$ from $\tilde I_{kj}$. Actually, these
multipliers just are to be found. Normalize them by fixing their triangular
matrix structure
\begin{eqnarray}
\Psi_{k,2m}(0,\nu)=\pmatrix{*&0\cr *&*},\quad
\Psi_{k,2m+1}(0,\nu)=\pmatrix{*&*\cr 0&*}.
                           \label{Psikj}
\end{eqnarray}
Comparison of these eigenfunctions with those of the $x$-equation gives rise
to the so-called gluing up equations:
\begin{eqnarray}
&\psi(0,\xi)\delta_{kj}(\xi)=
T(\xi)\Psi_{k,j}(0,\nu)\Delta_{kj}(\nu)&
\label{123}\\
&\mbox{for}\, j=1,2,3; \,k=1,2;&\nonumber\\
&\phi(0,\xi)\delta_{kj}(\xi)=
T(\xi)\Psi_{kj}(0,\nu)\Delta_{kj}(\nu)&
\label{456}\\
&\mbox{for}\, j=4,5,6; \,k=1,2.&\nonumber
\end{eqnarray}
The left side of these equations gives an expression of arbitrary solution of
the scattering problem for the $x$-equation while the right side gives that of
the other one. Here the right side multipliers $\delta_{kj}(\xi)$ are upper
(lower) triangular matrix-valued functions defined on $I_{kj}$ for $j=1,2,3$
($j=4,5,6,$ respectively) and $\Delta_{kj}(\nu)$ are upper (lower) triangular
for $j=2,4,6$ ($j=1,3.5$) matrices defined on $\tilde I_{kj}$. Thus two
multipliers $\delta_{kj}$ and $\Delta_{kj}$ are of one and the same matrix
structure for only four choices of subindices $(k,j)$:  $(1,2)$,  $(2,2)$,
$(5,1)$,  $(5,2)$. In these cases the equations (\ref{123}) and (\ref{456})
are effectively solved by reducing to the problem of triangular factorisation
of matrices. For example, when $k=1$, $j=2$ one has
$$
T^{-1}(\xi)
\psi(0,\xi)
=\Psi_{1,2}(0,\nu(\xi))\Delta_{1,2}(\nu(\xi))
\delta^{-1}_{1,2}(\xi),\quad \xi\in I_{1,2}.
$$
The rest of functions are defined by imposing the following
involution
\bb
\Psi_{kj}(\nu(\xi))=\Psi_{k,j+2}(\nu(\omega\xi))=
\Psi_{k,j+4}(\nu(\omega^2\xi)),
\label{inv1}\ee
which is completely consistent with the equations
(\ref{123}), (\ref{456}) as well as the involutions
\bb
\Psi_{kj}(-\nu(-\xi))=\sigma_1\Psi_{mn}(\nu(\xi))\sigma_1
\label{inv2}\ee
and
\bb
\overline{\Psi_{kj}(-\nu(-\xi))}=\Psi_{mn}(\bar\nu(\bar\xi)),
\label{inv3}\ee
where the line over a letter means the usual complex conjugation.

Having all of the functions $\Psi_{kj}(t,\nu)$ at the point
$t=0$ one can find the conjugation matrices of the
Riemann-Hilbert-Carleman problem on the Riemann surface $H$
with the contour of jumps $\tilde L=\nu(L)$
associated with the $t$-equation from the
Lax pair:
\bb
\Psi_{kj}(t,\nu)=\Psi_{mn}(t,\nu)R_{kjmn}(t,\nu).
\label{nurp}\ee
The conjugation matrices explicitly depend on $t$:
\bb
R_{kjmn}(t,\nu)=e^{-4i\xi t\sigma_3}R_{jkmn}(0,\nu)e^{4i\xi
t\sigma_3}
\label{rt}\ee
By solving the Riemann problem one finds the eigenfunctions $\Psi_{kj}(t,\nu)$
for all $t>0$ and recovers the linear system (\ref{ytn}), i.e. its unknown
coefficient $u_x(0,t)$. Afterwards the equation for the time evolution of the
scattering matrix (\ref{st}) becomes linear. On the other hand side the
scattering matrix $s(t,\xi)$ can directly be expressed in terms of the
functions $s(0,\xi)$ and $\Psi_{kj}(t,\nu).$ And the last step consisting of
recovering the potential $u(x,t)$ for a given $s(\xi,t)$ called the inverse
scattering problem is studied very much (see, for instance, \cite{zah}).

To conclude we remark that since the spectrum of $t$-equation
for $x=0$ lies on a two genus Riemann surface then it is
natural to expect that $u_x(0,t)$ has finite-gap behaviour
at infinity.


\section{Acknowledgments}

This work has been supported by the Russian Foundation for Basic
Research (grants \# 99-01-00431 and \# 98-01-00576).

\vspace{2cm}
\pagebreak
}

\begin{figure}[h]            \label{fig:pattern1}
\setlength{\unitlength}{0.05em}
\begin{picture}(200,410)(-350, -215)
 \put(-19, -30){\line( 2,3){36}} %
 \put(-102, -149){\line( 2,3){81}}
 \put(-98, -152){\line( 2,3){81}}

\put( 13,21){$\bullet$}
 \put(-21,21){$\bullet$}
 \put( 14,-32){$\bullet$}
 \put(27,-5){$\bullet$}
 \put(-37,-5){$\bullet$}
 \put(-22,-32){$\bullet$}

 \put(19, 24){\line( 2,3){85}}
 \put(16, 27){\line( 2,3){85}}
 \put(22, 23){\line(-1,1){12}}
 \put(-22, 23){\line(1,1){12}}
 \put(22,-23){\line(1,-1){12}}
 \put(-22, -23){\line(-1,-1){12}}

 \put( 220,-10){$\ell_1$}
 \put(-240, -10 ){$\ell_4$}

 \put( 17,-26){\line(-2,3){33}} %
 \put( 103, -152){\line(-2,3){83}}
 \put( 100, -156){\line(-2,3){85}}

 \put( -19, 23){\line(-2,3){85}}
 \put( -16, 26){\line(-2,3){85}}

 \put( 110,165){$\ell_2$}
 \put(-120, -175){$\ell_5$}

 \put(-35,   -1){\line( 1,0){64}} %
 \put(-200,    1){\line( 1,0){167}}
 \put(-200,   -4){\line( 1,0){167}}

 \put(32,    1){\line( 1,0){150}}
 \put(32,    -4){\line( 1,0){150}}

 \put(-120,165){$\ell_3$}
 \put( 110, -175){$\ell_6$}
 \put( 140,  70 ){$I_1$} \put( 140, -85 ){$I_6$}
 \put(-150,  70 ){$I_3$} \put(-150, -85 ){$I_4$}
 \put( -5 , 130 ){$I_2$} \put( -10, -140){$I_5$}
 \put(-140,-220 ){Pic.1. The double sheeted Riemann surface $\Gamma$.}
\end{picture}
\end{figure}
\begin{figure}[s]            \label{fig:pattern2}
\setlength{\unitlength}{0.05em}
\begin{picture}(200,410)(-350, -215)

 \put(-200,    0){\line( 1,0){400}}
 \put(-3,40){\line( 0,1){180}}
 \put(3,40){\line( 0,1){180}}
 \put(-3,-40){\line( 0,-1){180}}
 \put(3,-40){\line( 0,-1){180}}
 \put(0,-40){\line( 0,1){80}}

 \put(200,0){\vector( 1,0){10}}

 \put(-4,38){$\bullet$}
 \put(-4,-43){$\bullet$}

 \put(10,40){$\nu_{+}$}
 \put(10,-40){$\nu_{-}$}

 \put(-140,-245){Pic.2. The six sheeted Riemann surface $H$}

\end{picture}
\end{figure}

\end{document}